\documentclass[prl,twocolumn,10pt,showpacs,showkeys,preprintnumbers,amsmath,amssymb,unsortedaddress,aps]{revtex4-1}
\pdfoutput=1

\usepackage{graphicx}
\usepackage{amssymb}
\usepackage{amsmath}
\usepackage{textcomp}

\graphicspath{ {./figures/} }

\newcommand{\zc}{\overline{z}}
\newcommand{\wc}{\overline{w}}

\newcommand{\uRe}{\operatorname{Re}}
\newcommand{\uIm}{\operatorname{Im}}
\newcommand{\uMs}{M_{\mathrm{S}}}
\newcommand{\umx}{m_{\mathrm{X}}}
\newcommand{\umy}{m_{\mathrm{Y}}}
\newcommand{\umz}{m_{\mathrm{Z}}}

\begin{document}

\title{Metastable states of sub-micron scale ferromagnetic periodic antidot arrays.}

\author{Andrei B. Bogatyr\"ev}
\affiliation{Institute for Numerical Mathematics, Russian Academy of Sciences, 8 Gubkina str., Moscow GSP-1, Russia 119991}
\email{ab.bogatyrev@gmail.com}

\author{Konstantin L. Metlov}
\affiliation{Donetsk Institute of Physics and Technology, 72 R. Luxembourg str., Donetsk, Ukraine 83114}
\email{metlov@fti.dn.ua}
\date{\today}

\begin{abstract}
The magnetic textures on nanoscale possess topological features due to the continuity of the magnetization vector field and its boundary conditions. In thin planar nanoelements, where the dependence of the magnetization across the thickness is inessential, the textures can be represented as a soup of 2-d topological solitons, corresponding to magnetic vortices and antivortices, which are the solutions of Skyrme's model. Topology of the element (of the boundary conditions) then imposes the restrictions on properties and locations of these objects. Periodic arrays of magnetic antidots have topology with infinite connectivity. In this work we classify and build an approximate analytical representation of metastable magnetization textures in such arrays and prove the conservation of their topological charge.
\end{abstract}
\maketitle

Today there is a renewed interest to the magnetic topological solitons (or skyrmions). They are ubiquitous and stable in thin films\cite{M61, M01_CT, M04} and in planar nanomagnets\cite{UP93,SOHSO00, MG02_JEMS}. Their defining feature is the presence of non-zero integer topological charge\cite{BP75}, which splits the magnetization textures into topological classes. There are (infinite in theory\cite{BP75}) energy barriers, separating the textures of different topological class in thin films and granting the topologically charged textures an enhanced stability.

There are also additional topological properties of these textures, resulting from interplay of the topology of the magnetization textures and the topology of the nanoelement itself, specifically its connectivity\cite{BM17}. In simply connected nanoelements the positions of vortices and antivortices are influenced only by weaker magnetostatic forces. Each additional degree of connectivity imposes an additional algebraic constraint on their positions\cite{BM17,Bogatyrev2017}, guarded by both the exchange and magnetostatic forces. Similar constraints arise in the case of periodic magnetization textures\cite{BM18}. They are a defining feature, making magnetic skyrmions different from magnetic bubbles\cite{BM18}, which are also stable, but can have their positions and chirality controlled independently.

\begin{figure}
\includegraphics[width=0.95\columnwidth]{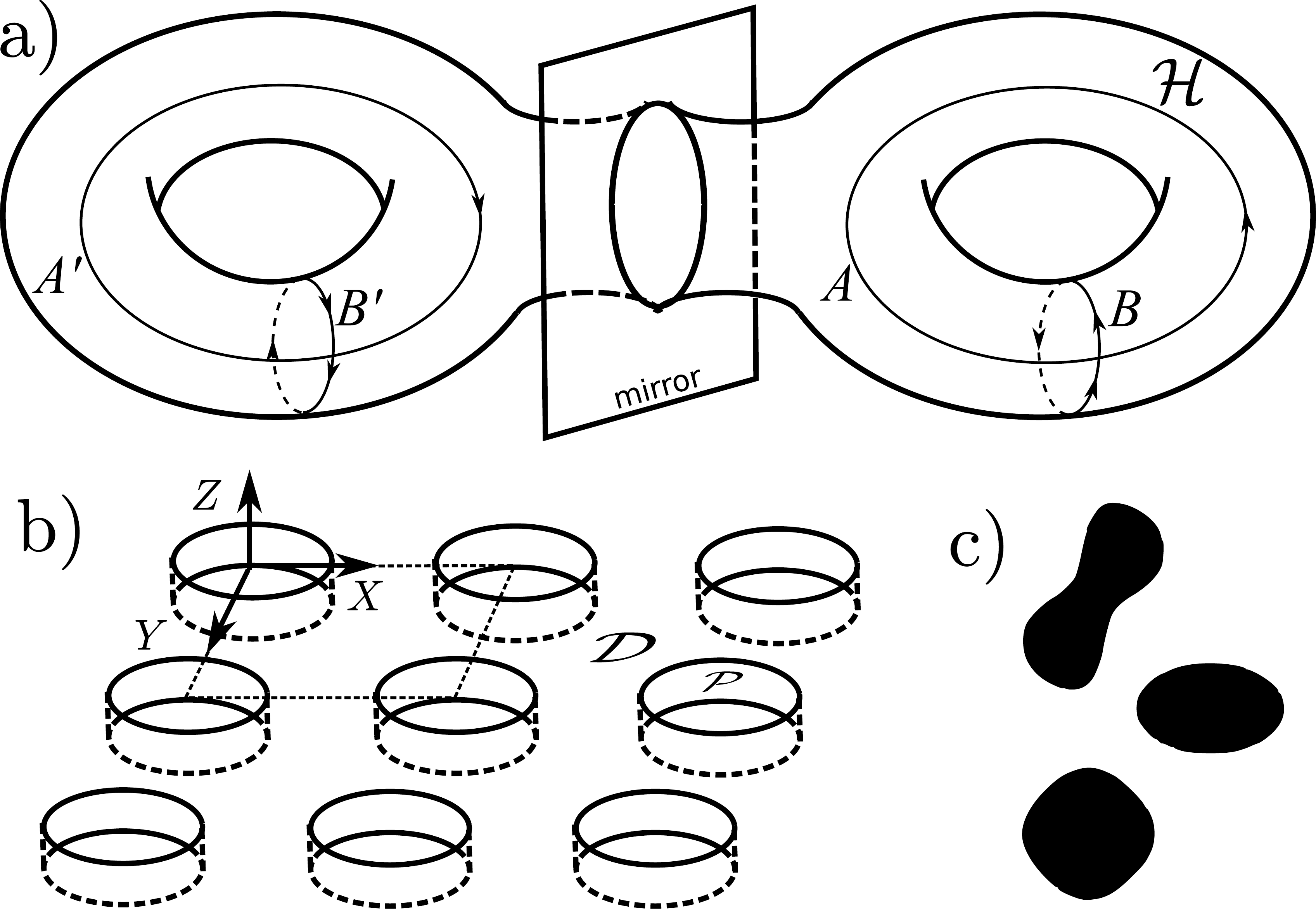}
\caption{\label{fig:Handle}a) Double $\cal DH$ of the handle $\cal H$ and basis of homologies on it: $A'=\sigma A$; $B'=-\sigma B$; b) an example of antidot lattice with unit cell outlined, ${\cal P}$ on the top face denotes one of the perforations and ${\cal D}$ is the top face of the whole antidot array; c) several ${\cal P}$ shapes, which can be parametrized by the complex linear fractional map.}
\end{figure}

Consideration of periodic antidot arrays\cite{CAB1997} is a logical next step, which we take in the present work. These arrays, shown schematically in Fig.~\ref{fig:Handle}b),
can be produced either lithographically\cite{CAB1997} or self-assembled\cite{ZGGBG2003}. Their magnetization textures were extensively studied experimentally and numerically\cite{YJSFM2000,TMGCANBG2010,Schneider2017}, but there is no analytical theory to date.

Formally, the antidot array is a thin film, containing a doubly periodic array of holes, shown in Fig ~\ref{fig:Handle}b) with the Cartesian coordinate system $\vec{r}=\{X, Y, Z\}$ such that the $X-Y$ plane coincides with the top face of the film. The holes are identical (but not necessarily circular) cylinders with the base ${\cal P}$ and their axis parallel to $Z$. Ferromagnetic material occupies a cylinder with the base ${\cal D}=\mathbb{C}\setminus({\cal P}+\mathbb{Z}+\tau\mathbb{Z})$, where $\mathbb{C}$ is the plane of complex variable $z=X+\imath Y$ and the base is invariant under the shifts 1, $\tau$, $Im~\tau>0$.

Every sufficiently small (but still macroscopic) volume inside the ferromagnet is uniformly magnetized\cite{AharoniBook}. The magnitude of this local magnetization vector $\vec{M}$ is a constant $\lvert\vec{M}\rvert=\uMs$ called the saturation magnetization, which is a material parameter. These vectors form a spatial distribution $\vec{M}(\vec{r})$ -- vector field, called the magnetization texture. If the film is sufficiently thin, the magnetization can be assumed to be constant across the film thickness $\partial \vec{M}/\partial Z = 0$. Then, without further loss of generality, the dimensionless magnetization vector $\vec{m}=\vec{M}/\uMs$ can be expressed via a complex function $w(z, \zc)$ using the stereographic projection
\begin{equation}
 \label{eq:magveccomp}
 \{\umx+\imath \umy, \umz \} = \frac{\{2 w(z,\zc), 1-w(z,\zc)\wc(z,\zc)\}}{1+w(z,\zc)\wc(z,\zc)},
\end{equation}
which guarantees $\lvert\vec{m}\rvert=1$. Here the line over the variable denotes complex conjugation $\zc=X -\imath Y$, and $\imath=\sqrt{-1}$.

In the absence of perforations, the magnetization textures of thin films are a subject of classical domains theory\cite{Hubert_Shafer} representing them as a set of uniformly magnetized domains, separated by the magnetic domain walls. The same can be expected if the distance between perforations is large or, if, instead of the infinite film, we deal with large thin film islands.

The smaller islands, which can not accommodate a domain (or even a domain wall), support a completely different set of magnetization textures, containing magnetic vortices\cite{UP93,SOHSO00,WWBPMW02}, which are 2-dimensional topological solitons or skyrmions. This special (``nanomagnetic'', since the relevant length scales are usually sub-micron) regime is much less studied. Here we will follow an approximate approach\cite{M01_solitons2, M10} assuming the existence of the well defined energy terms hierarchy and based on the sequential energy terms minimization. The exchange energy is dominating in sub-micron elements, because the positive magnetostatic self-energy of magnetic poles gets more and more compensated by their negative interaction energy as they are brought closer together when the magnet size decreases. Ultimately, the exchange wins completely and the magnetization becomes (quasi-)uniform, but before this happens there is a certain range of island sizes, where the isolated 2-d topological solitons can be directly observed\cite{UP93,SOHSO00,MG02_JEMS}. For multiply-connected regions\cite{BM15,BM17} (such as ring) the criterion becomes not the overall size of the element, but the characteristic size of the magnetic regions it contains (width of the ring). The same can be expected for infinitely-connected perforated thin films. It is very hard (if not impossible) to give the specific estimate of the size limits, because the long-range character of the magnetostatic interactions makes them strongly shape-dependent. In the following we shall just assume that the required energy hierarchy holds and see if the resulting theoretical magnetization textures correspond to the experimentally observed ones.

In the sequential minimization framework\cite{M10} the energy of face magnetic charges (on the top/bottom interfaces of the film) under the assumption of the magnetization vector field being locally extremal with respect to the exchange energy can be minimized by representing $w(z, \zc)$ as a soliton-meron\cite{G78} join
\begin{equation}
  \label{eq:sol_SM}
  w(z,\overline{z})=\left\{
    \begin{array}{ll}
      f(z)/c_1 & |f(z)| \leq c_1 \\
      f(z)/|f(z)| & c_1<|f(z)| \leq c_2\\
      f(z)/c_2 & |f(z)| > c_2
    \end{array}
    \right.,
\end{equation}
where $f(z)$ is an analytical function of complex variable and $c_1$,$c_2$ are free scalar parameters. Next, to zero-out the magnetostatic energy on the sides of the perforations $f(z)$ can be chosen as a solution of Riemann-Hilbert problem of finding the analytical in $\cal D$ complex function with no normal components at the boundary
\begin{equation}
   \label{eq:boundary}
   \left. \uRe f(z) \overline{n(z)}\right|_{z\in \partial{\cal D}} = 0,
\end{equation}
where $n(z)$ is the complex normal to the boundary of ${\cal D}$.

Let us now compute these magnetization textures for the case of the thin film, perforated with doubly-periodic lattice of holes, expressing  the solution in terms of real meromorphic differentials\cite{Bogatyrev2017} $d\xi:=dz/f(z)$ which automatically take into account boundary condition \eqref{eq:boundary}.

The factor of the perforated plane with exactly one hole in each elementary cell by rank 2 lattice of shifts is a torus with one hole (=handle) ${\cal H}$, shown in Fig.~\ref{fig:Handle}a). To describe all real meromorphic differentials on the handle we reflect the handle with respect to its boundary to get its Schottky double. The latter is a genus two surface without boundary, which admits the anticonformal involution (reflection) $\sigma$. As every genus two surface, the double is conformally equivalent to a hyperelliptic curve which in our case has an additional symmetry:
\begin{equation}
\label{AlgDH}
{\cal DH}:=\{(x,w)\in \mathbb{C}^2: ~w^2=\prod_{s=1}^3(x-x_s)(x-\overline{x_s}) \}
\end{equation}
with the branch points $x_1,x_2,x_3,$ strictly in the upper half plane. The reflection in this model acts as 
$\sigma(x,w):=(\bar{x},\bar{w})$ and its only real oval is the total lift of the extended real axis.
Each meromorphic differential on the surface \eqref{AlgDH} has an appearance
\begin{equation}
\label{Rdxi}
d\xi=(R_1(x)+w R_2(x))~dx,
\end{equation}
where $R_s(x)$ are rational functions (quotient of two polynomials). This becomes real differential (with the symmetry $\sigma d\xi=\overline{d\xi}$) exactly when $R_s(x)$ are {\em real} rational functions. It describes the space of lattice-invariant magnetization states for our model in case of one inclusion per unit cell.

To give explicit formulas for the states, fix a rank two lattice $L(\tau):=\mathbb{Z}+\tau\mathbb{Z}$,  $\uIm\tau>0$ on a complex plane of variable $z$ and consider the set $V$ of four points on the Riemann sphere
\begin{equation}
\label{V}
V:=\{0,\infty, -(\theta_0/\theta_3)^2, -(\theta_3/\theta_0)^2\}
\end{equation}
where $\theta_0$ and $\theta_3$ are so called theta constants, that is the values at $z=0$ of the appropriate standard (we use the definition from the book of Akhiezer\cite{akhiezer1990translation}) theta functions
$\theta_s(z|\tau)$ of modulus $\tau$. Those four points represent the branch points of a 2-sheeted model of the torus $\mathbb{C}/L(\tau)$.

Next step is to choose a complex linear fractional map $x=l(v)=(a_1v + a_2)/(a_3 v + a_4)$,  essentially depending on three complex constants,  such that the set $X:=\{x_s\}_{s=1,\dots,4}=l(V)$ contains just one point with strictly negative imaginary part (let it be $x_4$)
and the rest of points have strictly positive imaginary part. Practically, one can separate one of the points of $V$ by a circle
(not necessarily small) or a line (=big circle) from the rest of the points and then linear fractionally map this
circle to the real line, which becomes the boundary of the hole in the torus. 

Now consider a genus two curve (excluding the branch point with negative imaginary part) \eqref{AlgDH}
This is the double surface of a handle from our previous consideration. Real differentials on it have the appearance \eqref{Rdxi}
This differential can be pulled back to the physical plane of variable $z$ 
with the mapping
\begin{equation}
x(z):=l(\frac{\theta_1^2(z|\tau)}{\theta_2^2(z|\tau)}).
\end{equation}
The inclusions in the $z$ plane are surrounded by the lines with real value of $x(z)$ and correspond to the regions where $\uIm x(z)<0$, their shape and size depend on the selection of the linear fractional map parameters $a_k$ and $\tau$. Some of the possible shapes for $\tau=\imath$ are shown in Fig.~\ref{fig:Handle}c) with $\{a_k\}$ being (from top to bottom): $\{1,\imath,-\imath,1/8-\imath\}$, $\{1,\imath,-\imath,-1+2\sqrt{2}\imath\}$, $\{1,\imath,-\imath,-1+\sqrt{2}\imath\}$. 

\begin{figure}
\includegraphics[width=0.95\columnwidth]{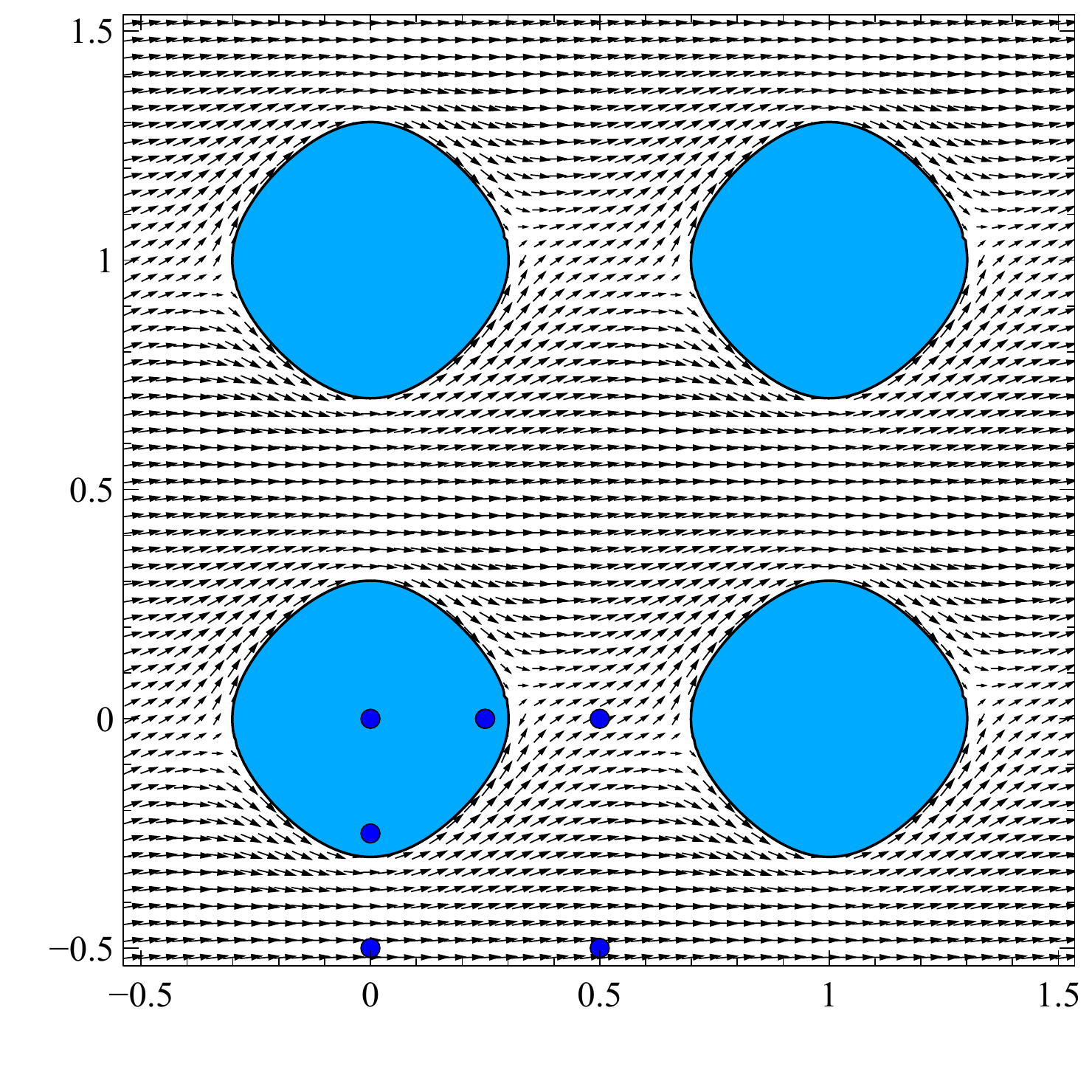}
\caption{\label{fig:groundstate} Ground state of the square array ($\tau=\imath$) of antidots with nearly circular holes ($\{a_k\}=\{1,\imath,-\imath,-1+\sqrt{2}\imath\}$), shown by shaded regions. The vectors are plotted from \eqref{eq:f}, \eqref{eq:sol_SM}, \eqref{eq:magveccomp} with $R_1(x)=0$, $1/R_2(x)=(x-(4\sqrt{2}+\imath)/11)(x-(4\sqrt{2}-\imath)/11)(x-(\sqrt{2}+\imath)/3)(x-(\sqrt{2}-\imath)/3)(x-\imath)(x+\imath)$ and $c_1=c_2=1/(4\pi)$. The dots show the poles of $R_2(x)$ mapped to one of the lattice cells in $z$.}
\end{figure}

It is important to choose the sign of the square root of the right hand side of \eqref{AlgDH} such that $w(x(z))$ is a single-valued meromorphic function in ${\cal D}$. Such sign selection is always possible, but can be tricky to achieve in practice. Finally, the solution of the Riemann-Hilbert problem \eqref{eq:boundary} is expressed as
\begin{equation}
\label{eq:f}
1/f(z)=(R_1(x(z))+w(x(z))R_2(x(z))) \frac{dx}{dz}(z) .
\end{equation}
Substituting this $f(z)$ into \eqref{eq:sol_SM} produces parametrized set of magnetization textures in the periodic antidot array with the elementary cell specified by $\tau$, and the shape of the inclusions given by the parameters $\{a_k\}$ of the linear fractional map. The functions $R_1(x)$ and $R_2(x)$ can be arbitrary rational functions with real coefficients. Different choice of the degrees of polynomials and their parameters correspond to different, topologically protected, metastable states of an antidot array. It is worth noting that the function  $w(x(z))$ itself has zeros and poles. Thus, to get access to simpler states it might be necessary to compensate them by choosing appropriate rational functions. This is necessary, in particular, to plot the state of the array, shown in Fig.~\ref{fig:groundstate},
which corresponds to the array's ground state, according to the experiments and simulations\cite{YJSFM2000,TMGCANBG2010,Schneider2017}. Many more metastable states are contained within \eqref{eq:f}. They can be the starting point for analytical consideration of statical, dynamical and spin-wave properties of sub-micron scale antidot arrays just like similar expressions were for simply-connected nanodots\cite{MG02_JEMS, GHKDB06, M13.dynamics}.

It may seem there is a little value in the above analytical formulas, because they are hard to evaluate (need special functions $\theta_s(z|\tau)$, need special procedure to compute the $w(x(z))$ with proper sign selection). These difficulties are not insurmountable, but the real power of analytical description comes from the possibility to make general statements about the magnetization textures without, actually, computing each and every one of them. For example, analyzing the expression \eqref{eq:f} like in \cite{BM17,Bogatyrev2017,BM18} it is possible to show that: 1) the number of vortexes in the elementary cell is one less than the number of antivortices; 2) the total number of quasiparticles on the boundary is integer (particles on the boundary are always counted with weight $1/2$).

Also, one can see in \eqref{eq:f} that the poles of $R_1$ and $R_2$, unless they are canceled, correspond to zeros of $f(z)$ -- or, to the vortex centers. Their location can be directly controlled by the coefficients in $R_1$ and $R_2$. But, because two terms (with $R_1$ and $R_2$) are summed in the right hand side of \eqref{eq:f}, the positions of zeros of this sum (antivortex centers) can not be independently controlled. This implies the existence of the (topological, because they are also mediated by the curve \eqref{AlgDH}) constraints, binding the positions of vortices and antivortices. Such constraints were already studied for the case of finite connectivity\cite{BM17,Bogatyrev2017} and for simply-connected periodic case\cite{BM18}, we plan to derive their explicit form in the forthcoming publication.

Now, looking again at \eqref{eq:f}, \eqref{Rdxi} or their counterpart in simply-connected case\cite{M10}, note that $f(z)$ is expressed in terms of rational functions of some complex variable $x$ with real coefficients; moreover, the boundary of the magnet corresponds to the line $\uIm x=0$. Since the roots of polynomials with real coefficients are either located directly on the axis $\uIm x=0$ or symmetrically around it, the real vortices/antivortices inside the material are compensated by ``imaginary'' vortices/antivortices outside. This is necessary to satisfy the boundary conditions \eqref{eq:boundary} and can be considered a variant of method of images. Since \eqref{eq:f} represents all the solutions of the linear boundary value problem for $f(z)$, it implies that vortices and antivortices can not cross the particle boundary (because their movement is symmetric, a vortex, exiting the medium, implies that the other vortex enters the medium at the same location, which is a contradiction). This proves the most important property of the considered model -- the conservation of the topological charge\cite{BP75}. Provided the boundary conditions \eqref{eq:boundary} hold, no vortex or antivortex can ever leave or enter the medium !

In the case of thin film\cite{BP75} the topological charge conservation is a property of an idealized continuous model. Magnetization textures of real films can change their topological charge via the vortex-antivortex pair nucleation or annihilation, accompanied by the formation of Bloch points (having infinite energy in continuous approximation, but finite in real magnets). This mechanism can still take place in bounded magnets. But also the change of the topological charge can happen via breaking \eqref{eq:boundary}, which is guarded by the magnetostatic forces (the magnetic poles on the boundary of the particle have strictly positive self-energy). Since the size of the magnetic solitons\cite{UP93} in soft ferromagnets with the exchange stiffness $C$ is of the order of the exchange length $L_E=\sqrt{C/(\mu_0 M_S^2)}$, it means that selecting the material with larger $C$ and $M_S$ allows to increase both the Bloch point energy and the penalty for violating the boundary condition \eqref{eq:boundary} without changing the scale of the solitons. This (or other ways of enforcing the boundary condition, such as placing the side of the magnetic islands/holes in contact with diamagnet or even a superconductor) may open the way to artificially create the medium, where two-dimensional magnetic topological solitons exist and their topological charge is strictly conserved in a wide range of metastable configurations. The additional constraints can subsequently be imposed by patterning the media, applying external magnetic field and/or locally modifying the material properties. Such systems can be instrumental for topological quantum computation or other means of information/signal processing using magnetic topological solitons.

To summarize: we have formulated an approximate analytical expressions for metastable magnetic states of sub-micron scale soft ferromagnetic periodic antidot arrays, proved the conservation of topological charge in this and simply-connected case.

\begin{acknowledgments}
The support of the Russian Foundation of Basic Research under the project {\tt RFBR 16-01-00568} is acknowledged. 
\end{acknowledgments}

\bibliography{bibliography}

\end{document}